\documentclass[a4paper,11pt]{article}
\usepackage{pos}

\usepackage{graphicx,array}
\usepackage{hyperref}
\usepackage{color}
\usepackage{amsmath,slashed,latexsym}
\usepackage{slashed}
\usepackage{latexsym}
\usepackage{bm}
\usepackage{braket}
\usepackage{placeins}
\usepackage{subcaption}
\usepackage{slashbox}

\title{Explaining Flavour Anomalies with Heavy Scalars}

\author*{Sokratis Trifinopoulos}

\affiliation{INFN, Sezione di Trieste, SISSA, Via Bonomea 265, 34136, Trieste, Italy}

\emailAdd{sokratis.trifinopoulos@ts.infn.it}

\abstract{Discrepancies between recent experimental results and their respective Standard Model predictions, known as flavour anomalies, are reported in semileptonic charged and neutral-current $B$-decays, the muon magnetic moment $(g-2)_\mu$, and the extraction of the Cabibbo angle. In this proceedings, we review two New Physics models that introduce two scalar mediators at the TeV scale and aim at a combined explanation of the flavour anomalies. The first model features the leptoquarks $S_1$ and $S_3$ and provides tree level solutions to both $B$-anomalies and one-loop level solution to the anomalous $(g-2)_\mu$. The second features the leptoquark $S_1$ and the charged singlet $\phi^+$. While $S_1$ provides the same solution to the charged-current $B$-anomaly and $(g-2)_\mu$ as in the first model, $\phi^+$ can accommodate the Cabibbo-angle anomaly independently and together with $S_1$ can resolve the neutral-current $B$-anomaly at one-loop.}

\FullConference{%
  Corfu Summer Institute 2021 "School and Workshops on Elementary Particle Physics and Gravity"\\
  29 August - 9 October 2021\\
  Corfu, Greece
}

\begin{document}
\maketitle

\section{Introduction}

Since a few years, the most significant experimental challenges to the Standard Model (SM) of particle physics have appeared in the field of flavour physics. In particular, the following four reported deviations have incited great interest and if univocally proven correct they would necessitate the postulation of a New Physics (NP) sector.

\begin{enumerate}
	\item \boldsymbol{$b \to c \tau \nu.$} A $3\sigma$ enhancement of the charged-current transition in $\tau$ vs. light leptons~\cite{Lees:2013uzd, Hirose:2016wfn, Aaij:2015yra, Aaij:2017deq, Abdesselam:2019dgh} with respect to the SM prediction~\cite{Bernlochner:2017jka,Bigi:2017jbd,Jaiswal:2017rve} manifests in the ratios $R_{D^{(*)}} = \frac{ \mathcal B(B \to D^{(*)} \tau \overline{\nu})}{ \mathcal B(B \to D^{(*)} \ell \overline{\nu} ) }$. This and the next deviation constitute hints of Lepton Flavor Universality (LFU) violation in semi-leptonic $B$-meson decays.
	\item \boldsymbol{$b \to s \ell \ell.$} A deficit of the neutral-current transition in muons vs. electrons~\cite{Aaij:2017vbb, Aaij:2019wad, Abdesselam:2019wac, Abdesselam:2019lab, Aaij:2021vac} is observed and encoded by the ratios $R_{K^{(*)}} = \frac{ \mathcal B(B \to K^{(*)} \mu\overline{\mu}) }{ \mathcal B(B \to K^{(*)} e \bar{e} )}$. The SM prediction is very accurate and equal to 1~\cite{Bordone:2016gaq}. Including also additional deviations reported in this channel the global significance amounts to $4.3 \sigma$~\cite{Isidori:2021vtc}.
	\item \boldsymbol{$(g-2)_{\mu}.$}
	The anomalous magnetic moment of the muon $a_\mu = (g-2)_{\mu}/2$ exhibits a long-standing deviation~\cite{Bennett:2006fi,Abi:2021gix}  which currently stands at an overall $4.2\sigma$ level~\cite{Aoyama:2020ynm}.
	\item \textbf{Cabibbo-Angle Anomaly (CAA).} The values of $V_{us}$ extracted from $K \to \pi \ell \nu$ decays, the ratio $\mathcal B(K \to \mu \nu)/\mathcal B(\pi \to \mu\nu)$ and CKM unitarity using the value of $V_{ud}$ estimated by superallowed nuclear $\beta$ decays do not coincide. The tension amounts to $3.6\sigma$ or $5.1\sigma$~\cite{Belfatto:2019swo,Grossman:2019bzp} depending on the input from the nuclear $\beta$ decays (i.e. Ref. \cite{Czarnecki:2019mwq}  or Ref. \cite{Seng:2018qru}).
\end{enumerate}

In this proceedings, we give a brief overview of two models based on scalar mediators that establish a connection between the flavour anomalies under the same LFU violating interpretation. After presenting the setup, we give the main results of the phenomenological analysis and finally comment on the merits and drawbacks of each scenario as well as their future prospects.

\section{Models}
\label{sec:models}

The two NP models under consideration are:
\begin{itemize}
	\item[$\diamond$] \boldsymbol{$S_1 + S_3$}~\cite{Crivellin:2017zlb,Crivellin:2019dwb,Saad:2020ihm,Gherardi:2020qhc,Lee:2021jdr,Marzocca:2021miv}: The model includes two scalar leptoquarks $S_1 = ({\bf \bar 3}, {\bf 1}, 1/3)$ and $S_3 = ({\bf \bar 3}, {\bf 3}, 1/3)$, where the quantum numbers under the SM gauge group $SU(3)_c \times SU(2)_L \times U(1)_Y$ are indicated. The interaction Lagrangian reads
\begin{equation}
	\mathcal L_{S_1 + S_3} = \left( (\lambda^{1L})_{i\alpha} \, \overline{q^c}_i  \,\epsilon\, \ell_\alpha
			+ (\lambda^{1R})_{i\alpha} \, \overline{u^c}_i   e_\alpha  \right) S_1 
			+ (\lambda^{3L})_{i\alpha} \, \overline{q^c}_i \,\epsilon\,  \sigma^I \ell_\alpha S_3^I + \rm{h.c.} ~,
	\label{eq:S1S3Model}
\end{equation}
where $\epsilon=i\sigma _2$. We denote SM quark and lepton fields by $q_i$, $u_i$, $d_i$, $\ell _\alpha$, and $e_\alpha$ and adopt latin letters ($i,\,j,\,k,\,\dots$) for quark flavor indices and greek letters ($\alpha,\,\beta,\,\gamma,\,\dots$) for lepton flavor indices. The weak-doublets quarks $q_i$ and leptons $\ell_\alpha$ are in the down-quark and charged-lepton  mass eigenstate bases.
	
	\item[$\diamond$] \boldsymbol{$S_1 + \phi^+$}~\cite{Marzocca:2021azj}: The particle content here contains $S_1$ and the singly charged scalar $\phi^+= ({\bf 1}, {\bf 1}, 1)$. The interaction Lagrangian reads	
\begin{equation}
\mathcal L_{S_1+\phi} = \frac{1}{2} \lambda_{\alpha \beta} \overline{\ell_{\alpha}^c} \epsilon \ell_{\beta} \phi^+ +\lambda_{i \alpha}^{1L} \overline{q_i^c} \epsilon \ell_{\alpha} S_1 + \lambda_{i \alpha}^{1R} \overline{u_i^c} e_{\alpha} S_1 + \rm{h.c.}~,
\label{eq:S1phiModel}
\end{equation}
Note that $SU(2)_L$ invariance enforces antisymmetry of the $\phi^+$ couplings to leptons: $\lambda_{\alpha \beta} = - \lambda_{\beta \alpha}$. \par
\end{itemize}

\begin{figure}[t]
\centering
  \begin{subfigure}{.25\textwidth}
    \includegraphics[width=1\linewidth]{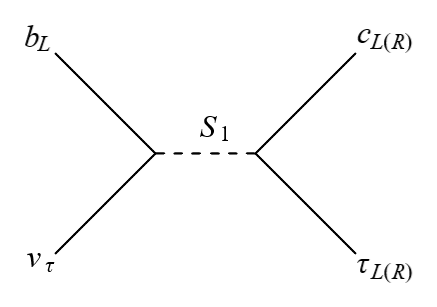}
    \caption{\label{fig:bctaunu}}
  \end{subfigure}%
	 \begin{subfigure}{.25\textwidth}
	   \includegraphics[width=1\linewidth]{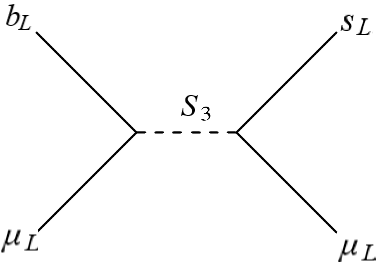}
    \caption{\label{fig:bctaunu_S3}}
  \end{subfigure}%
  \begin{subfigure}{.25\textwidth}
    \includegraphics[width=1\linewidth]{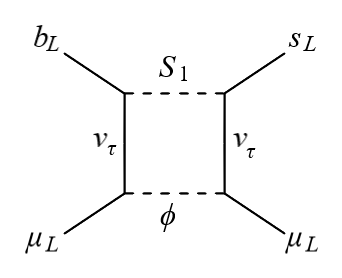}
    \caption{\label{fig:bsmumu}}
  \end{subfigure}
    \begin{subfigure}{.25\textwidth}
    \includegraphics[width=1\linewidth]{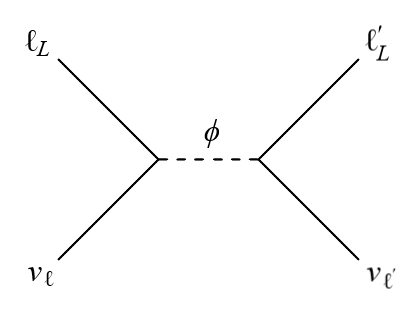}
    \caption{\label{fig:llpnunu_tree}}
  \end{subfigure}%
    \begin{subfigure}{.25\textwidth}
    \includegraphics[width=1\linewidth]{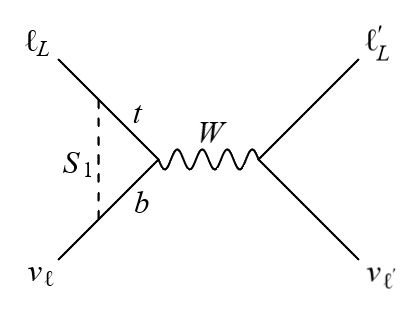}
    \caption{\label{fig:llpnunu_loop}}
  \end{subfigure}
      \begin{subfigure}{.25\textwidth}
    \includegraphics[width=1\linewidth]{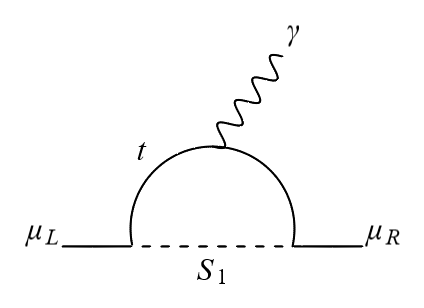}
    \caption{\label{fig:g-2}}
  \end{subfigure}
\caption{The diagrams that generate the dominant contributions to the flavour anomalies.}
\label{fig:anomalies_diag}
\end{figure}

The diagrams that yield the leading contributions to the anomalous observables can be found in Fig.~\ref{fig:anomalies_diag}. For explicit expressions of these contributions (as well as to all other relevant observables) for the $S_1 + S_3$ model we refer to Appendix A in Ref.~\cite{Gherardi:2020qhc} and for the $S_1 + \phi$ to Sec. 3 and the Appendix in Ref.~\cite{Marzocca:2021azj}. In Table \ref{tab:NP_couplings} we list the relevant NP couplings for each anomaly that are employed in the numerical analysis. For the case of $S_1 + S_3$ and $S_1 + \phi$ the couplings $\lambda_{t\tau}^{1R}$ and $\lambda_{c\mu}^{1R}$, respectively, are also required to cancel an otherwise excessive contribution to $\tau \to \mu \gamma$. The rest of the couplings are assumed to be negligible because they play no role in the fit of the anomalies, e.g. the leptoquark couplings to the first generation are set to zero because the observed deviations in $B$-decays involve primarily the second and third generations and also due to the strong constraints on $s\leftrightarrow d$ transitions. It can be shown that this approximation is stable under renormalization group effects.

\begin{table}[t]
\centering 
{\setlength\extrarowheight{8pt}%
\begin{tabular}{|c|c|c|}
\hline 
\backslashbox{Anomaly}{Model}  & $S_1+S_3$ & $S_1+\phi$ \tabularnewline
\hline 
\hline 
$b \to c \tau \nu$ &  $\lambda_{[b,s]\tau}^{1L},\lambda_{c\tau}^{1R},\lambda_{[b,s]\tau}^{3L},$ & $\lambda_{b\tau}^{1L},\lambda_{c\tau}^{1R}$ \\
$b \to s \ell \ell$ & $\lambda_{[b,s]\mu}^{3L}$ & $\lambda_{\mu\tau},\lambda_{b[\tau,\mu]}^{1L},\lambda_{s\tau}^{1L},\lambda_{c\mu}^{1R}$ \\
$(g-2)_\mu$ & $\lambda_{b\mu}^{1L}, \lambda_{t\mu}^{1R}$ &  $\lambda_{b\mu}^{1L}, \lambda_{t\mu}^{1R}$\\
CAA & - & $\lambda_{e\mu}, \lambda_{b\mu}^{1L}$ \\
\hline 
\end{tabular}}

\caption{\label{tab:NP_couplings}Summary of the couplings relevant to the flavour anomalies for each model.}
\end{table}

\section{Results and Conclusions}
\label{sec:Conclusions}

We build a global likelihood $\chi^2$ with all the relevant observables (see Table 2 and 3 in Ref.~\cite{Gherardi:2020qhc} and Table I in Ref.~\cite{Marzocca:2021azj}) and find the best-fit point by minimizing the $\chi^2$. For the $S_1+S_3$ model the fit prefers masses close to the current experimental bounds around $1~\rm TeV$, while for the $S_1+\phi$ the masses are larger, namely $5.5~\rm TeV$. This is due to the fact that the contributions from the two models to $b \to s \mu\mu$ scale differently, since the one is a tree-level effect and the other a loop-level.

\begin{figure*}[t]
\centering
\includegraphics[height=5.5cm]{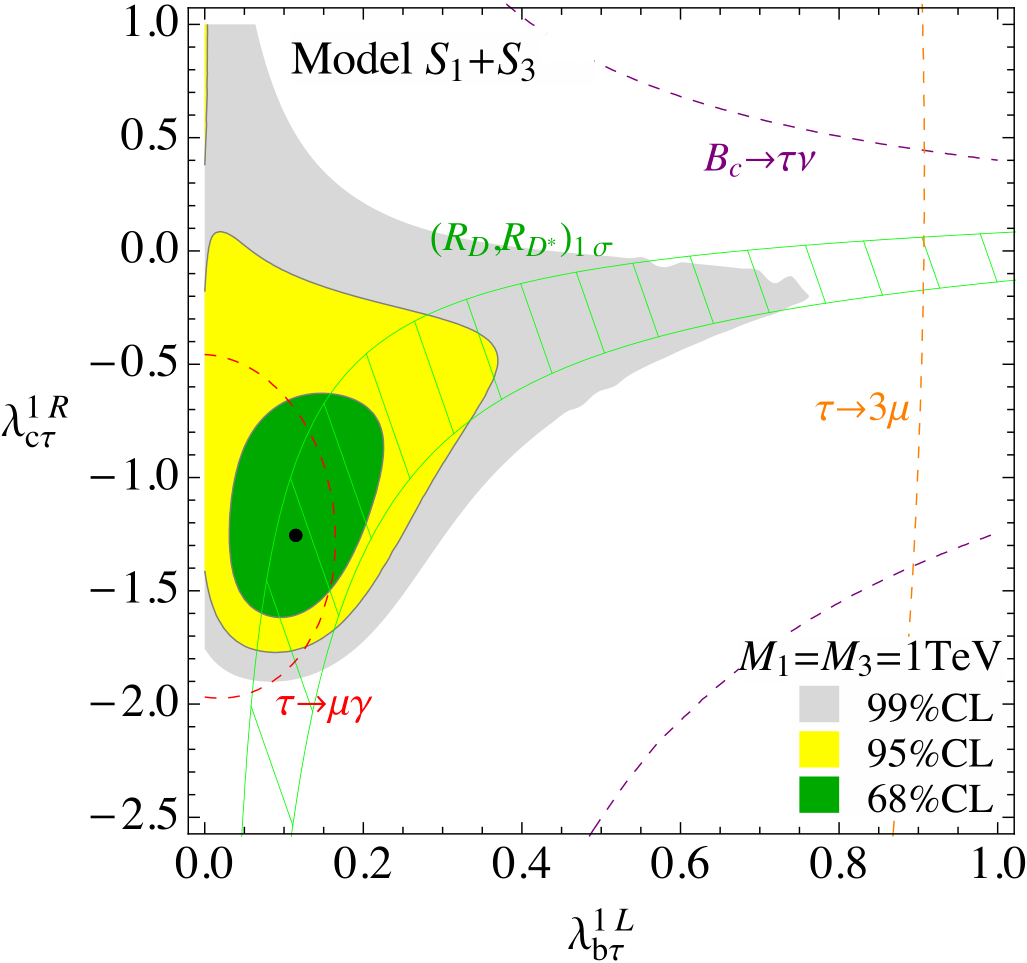} 
\includegraphics[height=5.5cm]{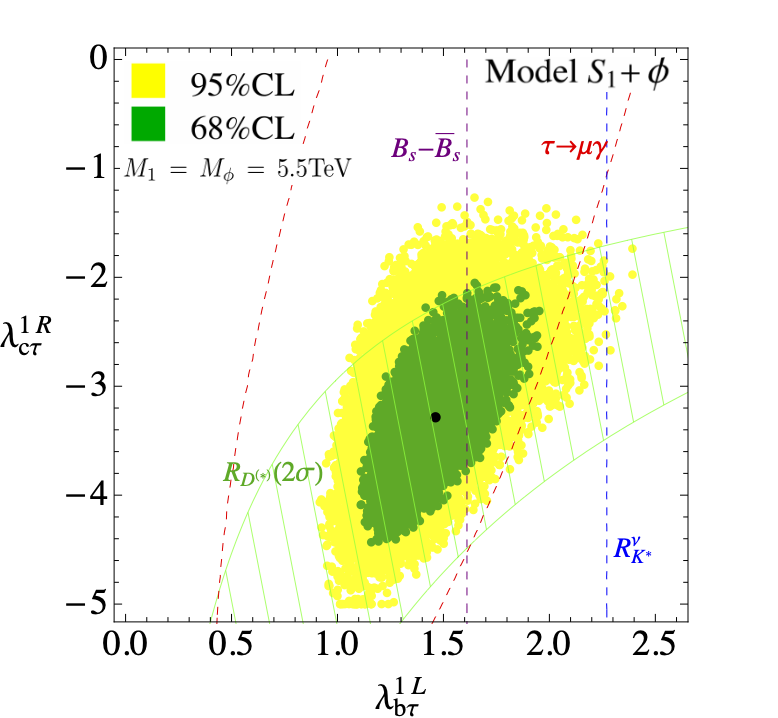} \\[0.5cm]
\includegraphics[height=5.5cm]{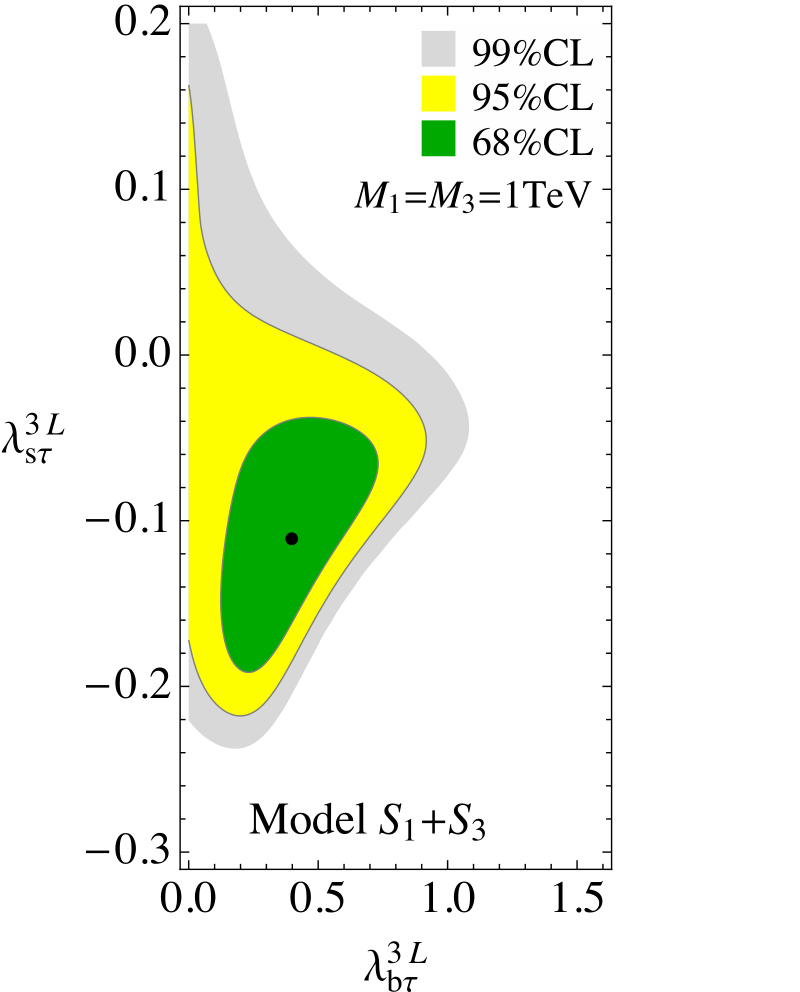} 
\includegraphics[height=5.5cm]{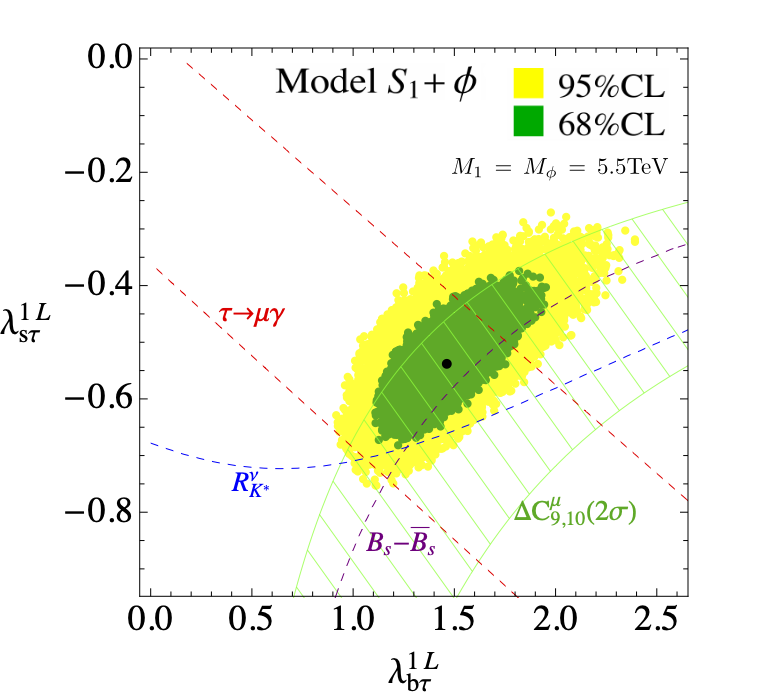} \\[0.5cm]
\includegraphics[height=5.5cm]{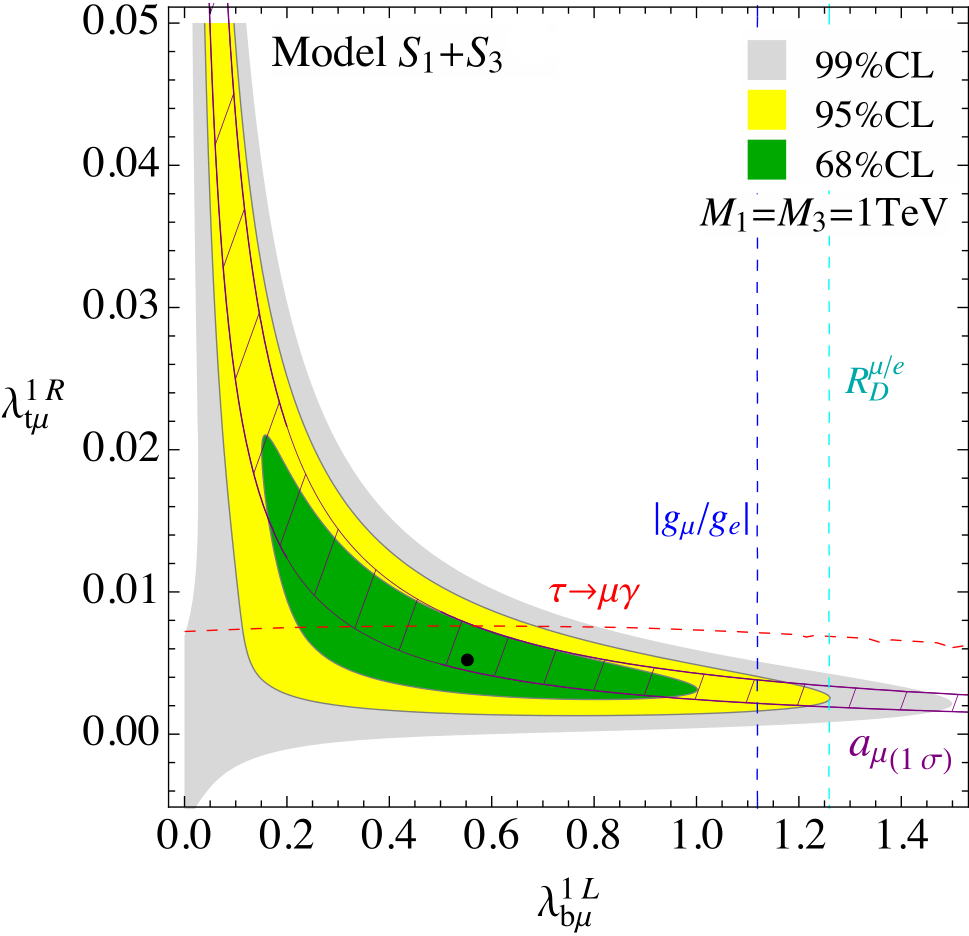} 
\includegraphics[height=5.5cm]{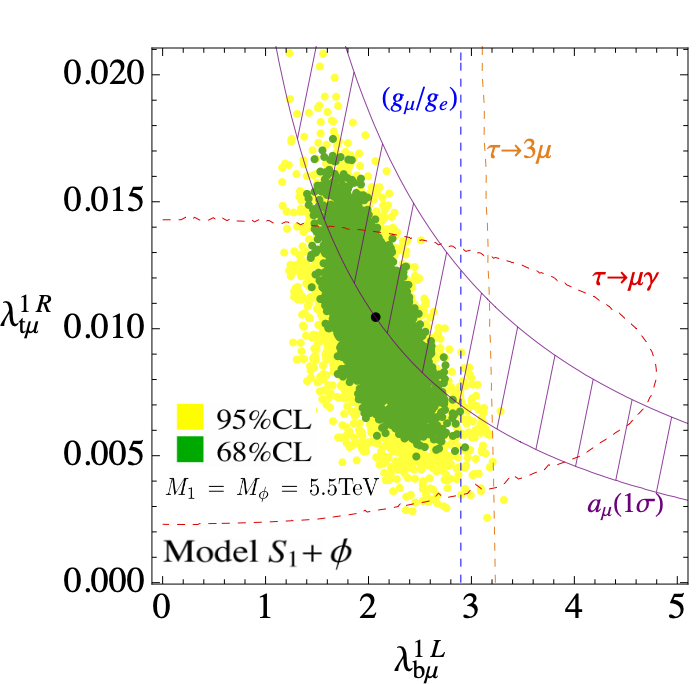} 
\caption{\label{fig:scan} Results of the parameter scan of the parameters for the two models, $S_1+S_3$ (left) and $S_1+\phi$ (right). The green (yellow) points are within $1\sigma$ ($2\sigma$) of the best-fit point, shown in black.}
\end{figure*}

Furthermore, in Fig. \ref{fig:scan} we show the results of a numerical scan on the parameter space of each model. Evidently, the two models are able to address at the $1\sigma$ level the discrepancies even after taking all the experimental constraints into consideration. They are among the very few models in the bibliography so far that can achieve this with less than two new particles introduced at the TeV scale. In fact, while most of the focus has been put on the $B$-anomalies and the $(g-2)_\mu$, the $S_1+\phi$ model associates also the CAA with them and in that respect, it is the most minimal combined explanation.

It is important to notice, that even though the scalar models are ultraviolet complete, we find that the couplings develop Landau poles at low scales, e.g. in the case of the $S_1+\phi$ around $20~\rm TeV$, and thus new degrees of freedom are expected to appear before those scales in order to regulate this effect. This is a generic issue of models trying to address the anomalies and employ rather sizeable Yukawa-type couplings \footnote{For this reason, the high-energy massless limit of the perturbativity analysis of Ref.~\cite{Allwicher:2021rtd} may not be a good approximation in the case where the mediator mass and the cut-off of the theory are separated by less than an order of magnitude.} and may point towards a scenario of composite scalars (see e.g. Ref.~\cite{Marzocca:2018wcf}). 

Moreover, it is a non-trivial task to find a flavour symmetry rationale for the flavour textures assumed in Table \ref{tab:NP_couplings}. In the case of $S_1+S_3$, approximate $U(2)$ flavour symmetries have been considered as a framework for accommodating the $B$-anomalies~\cite{Gherardi:2020qhc}. However, the right-handed couplings required for $(g-2)_\mu$ are highly suppressed in this case. Most importantly, a recent analysis~\cite{Marzocca:2021miv} has shown that in this case the bound of $\mathcal B (K_L \to \mu^+\mu^-)$ from NA62~\cite{NA62:2021zjw} already puts the model in tension. We leave the investigation of a more encompassing and viable flavour hypothesis for future works.

In the near future both models will be probed, since the LHCb and Belle-II experiments will reach a verdict on the nature of the $B$-anomalies, while the Fermilab $(g-2)_\mu$ experiment is expected to reduce further the experimental uncertainty. Regarding the CAA, experimental developments are expected in the existing precision observables used for the determination of the Cabibbo angle. Finally, among the most important model-specific signatures are for both models, sizeable effects in $B_s$ mixing and $B\to K^{(*)}\bar \nu\nu$; for $S_1+S_3$ the channels $b \to s \tau \tau$ and $b \to s \tau \mu$; and for $S_1+\phi$, the channel $b \to s \mu e$ and various lepton-flavour violating $\tau$ decays.

\FloatBarrier

\section*{Acknowledgments}

The author expresses sincere thanks to the organisers for the invitation and their hospitality and acknowledges the grant MIUR contract 2017L5W2PT.

\bibliographystyle{JHEP}
	
\bibliography{draft}

\end{document}